\documentclass[%
reprint,
superscriptaddress,
amsmath,
amssymb,
aps,
]{revtex4-2}

\usepackage{graphicx}
\usepackage{dcolumn}
\usepackage{bm}
\usepackage{hyperref}
\usepackage{upgreek}
\usepackage{multirow}

\begin{document}

\preprint{AIP/123-QED}

\title{Jitter in photon-number-resolved detection by superconducting nanowires}

\author{Mariia Sidorova}
\email{mariia.sidorova@dlr.de}
\affiliation{Humboldt-Universität zu Berlin, Department of Physics, Newtonstr. 15, 12489 Berlin, Germany}
\affiliation{German Aerospace Center (DLR), Institute of Optical Sensor Systems, Rutherfordstr. 2, 12489, Berlin, Germany}

\author{Timon Schapeler}
\affiliation{Department of Physics, Paderborn University, Warburger Str. 100, 33098 Paderborn, Germany}
\affiliation{Institute for Photonic Quantum Systems (PhoQS), Paderborn Unviersity, Warburger Str. 100, 33098, Paderborn, Germany}

\author{Alexej D. Semenov}
\affiliation{German Aerospace Center (DLR), Institute of Optical Sensor Systems, Rutherfordstr. 2, 12489, Berlin, Germany}

\author{Fabian Schlue}
\affiliation{Integrated Quantum Optics Group, Institute for Photonic Quantum Systems (PhoQS), Paderborn University, Warburger Str. 100, 33098 Paderborn, Germany}

\author{Michael Stefszky}
\affiliation{Integrated Quantum Optics Group, Institute for Photonic Quantum Systems (PhoQS), Paderborn University, Warburger Str. 100, 33098 Paderborn, Germany}

\author{Benjamin Brecht}
\affiliation{Integrated Quantum Optics Group, Institute for Photonic Quantum Systems (PhoQS), Paderborn University, Warburger Str. 100, 33098 Paderborn, Germany}

\author{Christine Silberhorn}
\affiliation{Integrated Quantum Optics Group, Institute for Photonic Quantum Systems (PhoQS), Paderborn University, Warburger Str. 100, 33098 Paderborn, Germany}

\author{Tim J. Bartley}
\affiliation{Department of Physics, Paderborn University, Warburger Str. 100, 33098 Paderborn, Germany}
\affiliation{Institute for Photonic Quantum Systems (PhoQS), Paderborn Unviersity, Warburger Str. 100, 33098, Paderborn, Germany}

\begin{abstract} 
By analyzing the physics of multi-photon absorption in superconducting nanowire single-photon detectors (SNSPDs), we identify physical components of jitter. From this, we formulate a quantitative physical model of the multi-photon detector response which combines local detection mechanism and local fluctuations (hotspot formation and intrinsic jitter) with thermoelectric dynamics of resistive domains. Our model provides an excellent description of the arrival-time histogram of a commercial SNSPD across several orders of magnitude, both in arrival-time probability and across mean photon number. This is achieved with just three fitting parameters: the scaling of the mean arrival time of voltage response pulses, as well as the Gaussian and exponential jitter components. Our findings have important implications for photon-number-resolving detector design, as well as applications requiring low jitter such as light detection and ranging (LIDAR).
\end{abstract}

\maketitle

\vspace*{-12mm}

\section{Introduction}
Photon-number measurements are a cornerstone of experimental quantum optics and fundamental to our understanding of the quantized nature of light. They are used to measure photon statistics in order to classify quantum optical states~\cite{mandel1979sub}, perform projective measurements which herald the presence of particular nonclassical photonic states~\cite{hong1986experimental,cooper2013experimental,tiedau2019scalability,endo2023non}. Photon-number-resolved (PNR) measurements are also required for a broad range of applications in photonic quantum information processing, such as error identification in photonic quantum computing~\cite{kok2007linear,walmsley2023light} and quantum key distribution~\cite{gisin2002quantum}. 

To measure photon number, one can rely on devices with an intrinsic photon-energy resolution. One archetypal example is the superconducting transition-edge sensor~\cite{lita2008counting,von2019quantum}. 
Nevertheless, these devices require significant millikelvin cryogenic overhead and their timing characteristics (on the nano- to microsecond level) limit the range of experiments and applications. A more experimentally accessible alternative has been to use spatially and/or temporally multiplexed threshold detectors~\cite{paul1996photon,fitch2003photon,achilles2003fiber}, also called ``on-off'' or ``click'' detectors, such as single-photon avalanche photodiodes (SPADs) or SNSPDs. These detectors trigger on detecting at least one photon. By splitting an unknown number of photons into multiple modes, each of which is incident on its own detector, one can achieve quasi-photon-number resolution. This approaches perfect PNR when the number of detectors is much larger than the number of photons to be measured~\cite{sperling2012true,jonsson2019evaluating}.

SNSPDs are now well-established as the detector of choice for many experiments and applications of quantum optics and photonics, because of their high efficiency, low noise, and accurate timing characteristics. SNSPDs were initially regarded as threshold detectors. However, Cahall et al.~\cite{cahall2017multi} observed that some degree of photon-number information could be obtained by analyzing the signal trace of the detector response, and since then a number of works have further refined this technique~\cite{zhu2020resolving,endo2021quantum,sempere-llagostera2022reducing,davis2022improved,sauer2023resolving,schapeler2024electrical,kong2024large,los2024high}. All of these rely on photons with an optical pulse duration shorter than the jitter of the detectors, typically below $100~\mathrm{ps}$. This is, however, well within the range of quantum light sources generated by nonlinearities~\cite{slussarenko2019photonic}.

Schapeler et al.~\cite{schapeler2024electrical} showed that almost all photon-number information is contained within the rising edge of the signal trace of the detector, and that this is most accurately measured with a time-to-digital converter conditioned on a threshold. Central to extracting photon-number information is therefore the ability to measure this arrival time accurately, which requires an understanding of the underlying photon-number-dependent jitter mechanisms. However, while models for SNSPD jitter are well accepted~\cite{vodolazov2019minimal, allmaras2019intrinsic}, the effects of photon number on jitter are far less well understood. Moreover, studying multi-photon absorption effects not only enhances our understanding of the underlying device physics of SNSPDs, but also allows detectors to be designed to maximize this figure of merit, thereby unlocking further applications of this fundamental photonic technology.

In this paper, we present a physical model which accounts for the photon-number-dependent jitter of SNSPDs, and apply it to experimental data obtained from a commercial device operated over a broad range of photon fluxes. We obtain excellent agreement between theory and experiment over several orders of magnitude in the arrival-time probability histogram. Based on this understanding, we suggest how SNSPDs can be designed to optimize their photon-number-resolving power. 

\section{Superconducting nanowire detector physics and jitter mechanisms}
Photon detection in superconducting nanowires involves three sequential stages~\cite{vodolazov2017single}: (i) hotspot formation, (ii) resistive domain evolution, and (iii) electrical signal propagation. Each stage introduces delays and variations that jointly determine the total jitter. The arrival-time histogram of voltage response pulses in single-photon detection with SNSPDs is well described with an exponentially-modified Gaussian (EMG) distribution~\cite{sidorova2017physical}
\begin{widetext}
\begin{equation}\label{eqn:EMGdist}
    P_\text{EMG}(t, \sigma, \tau, \mu) = \frac{1}{2 \tau} \exp\left(\frac{1}{2 \tau} \left[\frac{\sigma^2}{\tau} - 2t  + 2\mu \right]\right) \left[ 1 - \text{erf}\left( \frac{\frac{\sigma^2}{\tau} - t + \mu}{\sigma \sqrt{2}} \right) \right].
\end{equation}
\end{widetext}
Here, $\mu$ is the mean value, and $\sigma$ and $\tau$ are the Gaussian and exponential components, respectively.

For SNSPDs, the total measured jitter is typically determined as the histogram width, given by the standard deviation $\sigma_\text{tot}=\sqrt{\sigma^2+\tau^2}$. The Gaussian jitter component can be decomposed into its constituent contributions
\begin{equation}\label{eqn:sigma}
\sigma = \sqrt{\sigma_{\text{noise}}^2 + \sigma_{\text{inst}}^2+\sigma_\textrm{opt}^2 + \sigma_{\text{geom}}^2 + \sigma_\textrm{int}^2}.
\end{equation}
Among these, the first three contributions are due to external sources. 
The effect of noise in the electronics manifests as jitter via the slew rate $\gamma_\text{slew}$, {i.e.} the interplay of underlying electrical noise, the finite rise time $t_\text{rise}$ and amplitude $A$ of the electrical pulse is given by $\sigma_{\text{noise}}=\sigma_\text{elec}/\gamma_\text{slew}=\sigma_\text{elec}\left(\text{d} A/\text{d} t\right)^{-1}\approx \sigma_\text{elec}t_\text{rise}/A$, assuming a linear slope over the rise time. Jitter from the data acquisition instrumentation (e.g. time-to-digital converter, trigger photodiode) is contained with $\sigma_\text{inst}$, while $\sigma_\textrm{opt}$ accounts for jitter arising from the optical pulse duration. Geometric jitter $\sigma_{\text{geom}}$ arises from the temporal uncertainty of where photons are absorbed along the wire, i.e., it is determined by the SNSPD geometry. The contribution $\sigma_\textrm{int}$ together with $\tau$ constitute the jitter intrinsic to the photon detection process, throughout the paper we refer to it as the intrinsic jitter ($\sqrt{\sigma_\textrm{int}^2 + \tau^2}$).

Intrinsic jitter is determined by the physics of the hotspot formation process. It arises from thermal fluctuations affecting the detection energy barrier that dominate other sources~\cite{semenov2020local} like Fano fluctuations (or phonon loss)~\cite{kozorezov2017fano}, position-dependent response~\cite{vodolazov2019minimal}, and non-uniformities in the superconducting energy gap or film thickness~\cite{carbillet2020spectroscopic}. After photon absorption, the hotspot region enters a metastable state with the free energy $U$ defined by the photon energy and the actual local temperature, which itself is the subject of background thermal fluctuations in the absence of photons. Thermal fluctuations in the hotspot~\cite{semenov2020local} (a Poisson stochastic process) may provide enough energy to escape the metastable state over the still remaining energy barrier $\Delta U$ (difference between the detection threshold and the actual energy). The escape rate (inverse first hitting time~\cite{lipton2018first}), distributed exponentially, varies with $\Delta U$. For large $\Delta U$, escapes are rare, producing an exponential tail in the arrival-time histogram parameterized by $\tau$, while for smaller $\Delta U$ (e.g., in high-energy photon detection), escapes occur frequently, and the histogram retains Gaussian shape inherent to the background fluctuations. The jitter in this case is defined by events driving the hotspot to energies above the detection threshold~\cite{vodolazov2019minimal}.

The arrival-time histogram in multi-photon detection can be described using a sum of $P_\text{EMG}$ (Eq.~\ref{eqn:EMGdist}) with different means $\mu$ corresponding to different photon numbers, each weighted by the probability for a given photon-number component in the underlying photon distribution $P_n$. For coherent light, this distribution is Poissonian with $P_n(\bar{n})=\frac{ \bar{n}^n e^{-\bar{n}}}{n!}$, resulting in
\begin{equation}\label{eqn:EMGsum}
    P_\text{PNR}(\bar{n}) = \sum_{n=1}^\infty P_n(\bar{n}) P_\text{EMG}(t, \sigma_n, \tau_n, \mu_n).
\end{equation}
The key challenge in generalizing this model for multi-photon absorption is therefore to account for photon-number dependence in each of the parameters in the $P_\textrm{EMG}$ distribution, namely 
$\sigma_n$, $\tau_n$ and $\mu_n$.
The photon-number dependence of $\mu_n$ has previously been studied by Nicolich et al.~\cite{nicolich2019universal}. Here, we refine this model and consider additionally the photon-number dependence of $\sigma$ and $\tau$ arising from multi-photon absorption.

To account for multi-photon effects in all these contributions, one must consider the underlying physics of different multi-photon absorption scenarios.

\begin{figure*}[th]
\centering\includegraphics[width=1\linewidth]{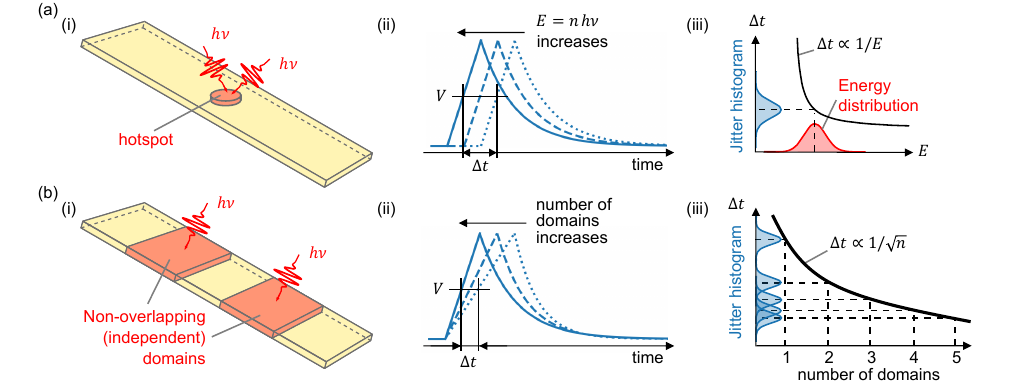}
\caption{Jitter in multi-photon detection. (a) Hotspot stage: (i) Upon multi-photon absorption at the same site, the energy deposited in the hotspot $E=n\cdot h\nu $ scales with the number of absorbed photons $n$, each of energy $h\nu$. (ii) Arrival time decreases as $E$ increase. (iii) Relation between deposited energy and jitter. Higher energy reduces the intrinsic delay time and narrows the jitter distribution. (b) Domain stage: (i) When $n$ photons generate $n$ non-overlapping resistive domains, (ii) the slope of the voltage response pulse becomes faster $\propto 1/\sqrt{n}$. (iii) shows the corresponding arrival-time histogram as a function of photon number.}\label{fig:jitterSchematic}
\end{figure*}

\subsection{Multi-photon absorption at the same site}\label{sec:sameSite}
When multiple photons ($n$), each of energy $h\nu$, are absorbed simultaneously within the hotspot size, the deposited energy is $E=n \cdot h\nu$ (see Fig.~\ref{fig:jitterSchematic}(a)(i)). Higher energy deposition leads to a shorter intrinsic delay, as this is proportional to $1/(E-E_0)$ due to the energy-dependent collapse time of the superconducting order parameter ($E_0$ is the photon energy corresponding to the detection threshold~\cite{sidorova2018timing,vodolazov2019minimal,allmaras2019intrinsic}), depicted in Fig.~\ref{fig:jitterSchematic}(a)(iii). Furthermore, for the same dispersion in the background energy fluctuations, intrinsic jitter decreases with $n$. Hence, for $E>E_0$, increasing $n$ results in earlier domain growth and, therefore, earlier voltage response pulse arrivals $\mu$ (alongside reduced intrinsic jitter). Note, this does not affect the dynamics of the resistive domain (i.e. rise time $t_\text{rise}$) but only the delay $\mu$ and the intrinsic jitter (see Fig.~\ref{fig:jitterSchematic}(a)(ii)). Despite the fact that for $E<E_0$ the detection mechanism is different and is governed by fluctuations~\cite{semenov2020local}, the qualitative result remains, i.e., intrinsic jitter decreases as $n$ increases. 

Given the small characteristic hotspot size ($L_T\approx50$~nm in NbTiN~\cite{sidorova2021magnetoconductance} and $L_T\approx220$~nm in WSi~\cite{sidorova2018nonbolometric}) the probability of multi-photon absorption is high in detectors with small active areas, such as those integrated with waveguides~\cite{jaha2024kinetic} or nano-detectors~\cite{renema2012modified}. If the area does not exceed the hotspot size, the intrinsic jitter monotonously decreases as both the number of photons and the mean photon number (light intensity) increase. 

For detectors with areas exceeding the hotspot size, there are several independent absorption sites $M_\mathrm{HS}$ over which incident photons are distributed randomly. This results in a combination of measurement events in which multiple photons are absorbed at the same site or independently at different sites. 

\subsection{Simultaneous multi-photon absorption at independent sites} \label{sec:independent}
When non-overlapping (independent) resistive domains (see Fig.~\ref{fig:jitterSchematic}(b)) are created, they share the same current, which correlates their dynamics. As the number of resistive domains increases, the nanowire resistance $R$ grows faster, causing the current to divert more rapidly into the external circuit and producing a voltage response pulse with a shorter rising time $t_{\mathrm{rise}}=L_k/R\propto 1/\sqrt{n}$ (see Fig.~\ref{fig:jitterSchematic}(b)). This is because $R \propto ntu$, where $u$ is the domain wall propagation rate, and the voltage rise $V(t)\propto (1-e^{-t^2nu})$ reaches $(1-e^{-1})$ when $t\propto 1/\sqrt{n}$. This results in photon-number dependent mean arrival times $\mu_n\propto1/\sqrt{n}$, as shown by Nicolich et al~\cite{nicolich2019universal}. The non-linear electrothermal model~\cite{sidorova2022phonon} refines this to $t_{\mathrm{rise}}\propto n^{-\alpha}$, where $\alpha\in\left[0.3,0.4\right]$ depending on detector parameters, in agreement with experimental results~\cite{kong2024large}. 

\subsection{Delayed mutli-photon absorption at independent sites}\label{sec:delayed}
When photons are absorbed at independent sites with a time delay, typically due to optical pulse durations $\sigma_\textrm{opt}$ longer than the hotspot formation time (about the electron-phonon cooling time), the detection process becomes more complex. The first photon creates a hotspot and a growing resistive domain, causing the current to start diverting into the external circuit. The second photon, absorbed at a different site, encounters a reduced current. At smaller currents, the remaining energy barrier is larger leading to even later emergence of the resistive domain and larger intrinsic jitter. The domain also grows more slowly under the reduced current, so that when the first resistive domain has already disappeared, the second domain may still exist. This increases the voltage rise time compared to the case of simultaneous photon absorption, affecting all parameters $\sigma_n,\tau_n,\mu_n$ in the model. As the delay between the photon absorption at independent sites increases, so does the jitter and the voltage rise time. When this delay becomes comparable to the resistive domain lifetime (approximately the electron cooling time to the substrate, $50$~ps for NbTiN~\cite{sidorova2024low}, $\sim800$~ps for WSi~\cite{sidorova2018nonbolometric} and $\sim70-130$~ps for MoSi (unpublished)), the second photon cannot be resolved. 

\subsection{Impact of Resistive Domain Overlap}\label{sec:overlap}
For sufficiently many independent absorption sites, each absorption and subsequent hotspot formation can be considered as independent events. While this limit is approached with large area detectors, in practice, overlapping domains compromise photon-number resolution, as two photons absorbed within the domain length $L_\mathrm{D}$ are detected as one. For a nanowire composed of $M_\mathrm{D}$ independent elements, each of length $L_\mathrm{D}$, the probability that no element contains more than one photon ($n \leq M_\mathrm{D}$, photons are uniformly distributed) is $P(\text{no overlap}) = \prod_{i=0}^{n-1} \frac{M_\mathrm{D}-i}{M_\mathrm{D}}$. This arises because the first photon can occupy any of the $M_\mathrm{D}$ elements with probability $1/M_\mathrm{D}$, the second photon avoids the already-occupied element with probability $(M_\mathrm{D}-1)/M_\mathrm{D}$, and so on. Then the probability that at least one element contains two photons is 
\begin{align}\label{eqn:Poverlap}
    P(\text{overlap}) &= 1 - P(\text{no overlap})\notag \\ 
    &= 1 - \prod_{i=0}^{n-1} (1-x_i) \approx 1 - \exp\left(-\frac{n(n-1)}{2M_\mathrm{D}}\right)\,,
\end{align}
which is valid for $x_i \ll 1$, where $x_i=i/M_\mathrm{D}$.

In principle, the probability of domain overlap as well as hotspot overlap is never zero, therefore, one should investigate the photon-number dependence of $\sigma_n$, $\tau_n$, and $\mu_n$.

\subsection{Multi-photon geometric jitter}\label{sec:geom}
Geometric jitter $\sigma_\textrm{geom}$ arises from spatial variations in the propagation time of the electrical signal from the photon absorption position $x$ to the readout electronics. If the nanowire can be treated as a transmission line (which may not hold under all circumstances~\cite{santavicca2016microwave}), two electrical pulses propagate with velocity $v$ towards either end of the nanowire with length $l$. In the differential readout scheme~\cite{calandri2016superconducting, kuzmin2019geometrical}, the absolute arrival time difference at the readout is $\Delta t = |x/v-(l-x)/v|=2x/v$. For a uniformly distributed probability of photon absorption, the distribution of $\Delta t/2=x/v$ is uniform and position-dependent and, therefore, causes jitter. In single-photon detection, the geometric jitter is $\sigma_{\mathrm{geom,1}} = l/(2\sqrt{3} v)$. For typical parameters of $v=6~\upmu$m/ps and $l=200$~$\upmu$m~\cite{calandri2016superconducting, kuzmin2019geometrical}, the single-photon geometric jitter is $\sigma_{\mathrm{geom,1}}\approx10$~ps.

\begin{figure*}[ht]
\centering\includegraphics[width=\textwidth]{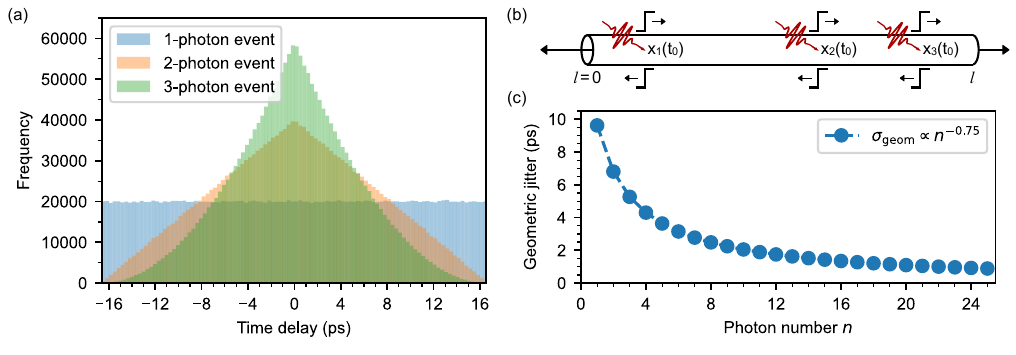}
\caption{(a) Distributions of time differences $\Delta t$ for one-, two-, and three-photon events show flat, triangular and peaked shapes, respectively. (b) Schematic showing the origin of geometric jitter. (c) Geometric jitter $\sigma_{\mathrm{geom}}$ as a function of photon number $n$.}\label{fig:geomJitter}
\end{figure*}

For two-photon detection, the distribution of the time difference $\Delta t$ follows a symmetric triangular distribution because at least one site is more likely to be closer to the nanowire ends (see orange bars in Fig.~\ref{fig:geomJitter}(a)). The geometric jitter for this case $\sigma_{\mathrm{geom,2}} = l/(2\sqrt{6} v)$ is reduced by a factor of $\sqrt{2}$ compared to single-photon detection. For three-photon detection, the time difference $\Delta t$ is determined by the two outermost absorption sites. Since one site always falls between the other two, the variation of $\Delta t$ is constrained, leading to a narrower, more peaked distribution than the triangular case for two sites (see Fig.~\ref{fig:geomJitter}(a)). Geometric jitter therefore decreases as the photon-number state increases, following $\sigma_{\textrm{geom},n}\propto n^{-{0.75}}$, (with the exponent determined numerically by Monte-Carlo simulations).

Now consider a conventional readout scheme, where one end of the nanowire is shorted. At the absorption site $x$, one electrical pulse propagates directly to the readout end, the other first reaches the shorted end, and then, propagating through the ground, arrives at the readout end. The arrival time of the “center of mass” $\Delta t = |t_1+t_2|/2=(l/v+l/v_g)/2$ is independent of position and thus does not cause jitter. Here, $v_g\approx140$~$\upmu$m/ps is the propagation velocity through the ground plane; since $v_g \gg v $ the contribution of the last term $l/v_g$ is negligible. However, geometric jitter may arise if the dispersion and losses differ between the nanowire and the ground plane~\cite{sidorova2017physical}. 

In experiments with coherent light, the observed distribution of arrival times results from a convolution of uniform, triangular, and peaked distributions, with their weights determined by the Poisson distribution of photon numbers in the coherent state. Exactly this was observed in Ref.~\cite{kuzmin2019geometrical} (see Fig.~3(a) therein), although it was not understood at that time.

\section{Comparison to experiment}
We verify our model (Eq.~\ref{eqn:EMGsum}) against experimental data from commercial SNSPDs, namely arrival-time histograms of the rising edge of an SNSPD trace. To fit the model to the data, we consider a relatively large area detector (thus neglecting multi-photon absorption at the same site and effects of resistive domain overlap) which gives rise to multi-photon geometric jitter. We also consider optical pulse durations much shorter than the hotspot formation time (neglecting delayed photon absorption in independent sites).

We thus seek the photon-number dependence of the parameters $\sigma_n$, $\tau_n$ and $\mu_n$ in Eq.~\ref{eqn:EMGsum}, and therefore the photon-number dependence on the constituents of $\sigma_n$ (Eq.~\ref{eqn:sigma}). Assuming no photon-number dependence arising from the instrumentation $\sigma_\text{inst}$ or optical pulse duration $\sigma_\text{opt}$, we are left with the intrinsic jitter $\sigma_{\text{int},n}$ and $\tau_n$, noise jitter $\sigma_{\text{noise},n}$, and geometric jitter $\sigma_{\text{geom},n}$
.  In principle, both $\sigma_\text{int}$ and $\tau$ have photon-number dependence, which will arise from the physics described above (fluctuations, detector geometry, material parameters, etc.). However, since we use a commercial device, we do not have access to this information, therefore we consider constant $\sigma_\text{int}$ and $\tau$ for all $n$ as fitting parameters. Further refinements to this simplified model to account for e.g. domain overlap would require further details of the SNSPD used (see Sec.~\ref{sec:overlap}), since there is always non-zero probability of domain overlap (Eq.~\ref{eqn:Poverlap}). Furthermore, we assume that the detector responds identically to incident optical pulses, such that the effects of time-walk~\cite{mueller2022time}, arising from count rate dependent variations in the bias current, can be neglected (operating at a sufficiently low repetition rate).
Under these assumptions, the photon-number dependence of the parameters are as follows:
\begin{subequations}\label{eqn:PNRdep}
\begin{align}
    \mu_n\rightarrow&~\frac{\Delta\mu}{\sqrt{n}}\propto\frac{t_{\text{rise},1}}{\sqrt{n}}\\
    \tau_n\rightarrow&~\tau\\
    \sigma_{\text{int},n}\rightarrow&~\sigma_\text{int}\\
    \sigma_{\text{noise},n}\rightarrow&~\frac{\sigma_\text{elec}}{\gamma_{\text{slew},1}\sqrt{n}}\approx\frac{\sigma_\text{elec}t_{\text{rise},1}}{A\sqrt{n}}\\
    \sigma_{\text{geom},n}\rightarrow&~\frac{\sigma_{\text{geom},1}}{n^{0.75}}
\end{align}
\end{subequations}
The variables $\Delta\mu$, $\tau$ and $\sigma_\text{int}$, are the three fitting parameters, while all other parameters can be empirically determined.

\begin{figure*}[ht]
    \centering
    \includegraphics[width=\linewidth]{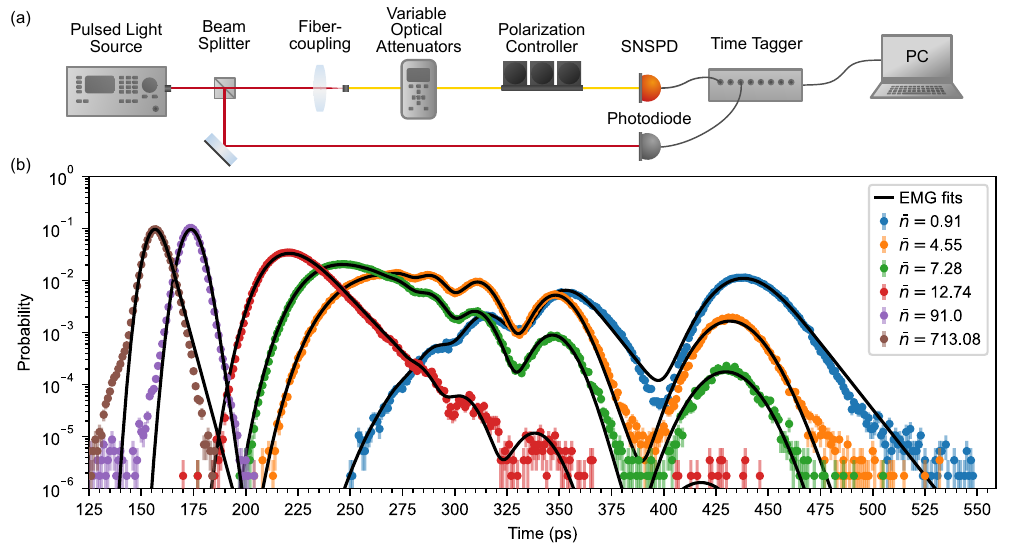}
    \caption{(a) Experimental setup to measure photon-number resolution with SNSPDs based on the rising edge arrival times relative to a trigger signal. Refer to the main text for more detail. (b) SNSPD rising edge arrival-time histograms for different mean photon numbers $\bar{n}$ specified in the legend (the error bars are based on Poisson errors of the counting statistics). The black lines correspond to the best fits based on a sum of exponentially-modified Gaussian distributions (Eq.~\ref{eqn:EMGsum}).}
    \label{fig:setup}
\end{figure*}

The experimental setup to acquire the arrival-time histograms is schematically shown in Fig.~\ref{fig:setup}(a). We use a pulsed light source at a wavelength of $1550~\mathrm{nm}$ with a repetition rate of $9.5~\mathrm{kHz}$ and 1~ps pulse duration. We split the optical signal with a beam splitter, where one part is directed to a fast photodiode (with $5~\mathrm{GHz}$ bandwidth) and the other part is fiber-coupled. Two variable optical attenuators are used to accurately set the mean photon number per pulse of the input states (this is done via a calibration process with uncertainty $\pm5\%$, see Ref.~\cite{schapeler2024electrical}). We optimize the polarization of the light with manual fiber polarization controllers to maximize the polarization-dependent detection efficiency of the SNSPD, specified by the manufacturer (Single Quantum) as $(91\pm3)\%$. The arrival times corresponding to crossing the rising and falling edge of the electrical signal of the SNSPD, as well as the trigger signal from the fast photodiode, are recorded using a Time Tagger X (Swabian Instruments, $<2~\mathrm{ps}$ std jitter). We choose the time tagger threshold at half the peak height of the SNSPD output trace, which gives optimal separation of photon-number events~\cite{schapeler2024electrical}.

The SNSPD is biased at $17.7~\upmu\mathrm{A}$ and has a switching current specified to be $19.3~\upmu\mathrm{A}$. The jitter of the SNSPD is specified by the manufacturer as $19~\mathrm{ps}$ FWHM (8~ps std). We assume that this value comprises all jitter components measured under single-photon illumination with optical pulses significantly shorter than the hotspot formation time (such that optical jitter may be neglected). The detector is read out across a $50~\Omega$ load resistor (single-ended readout), and then amplified.

We measure a total of $122$ different input states with varying mean photon numbers per pulse $\bar{n}$ (linearly increasing from $0$ to $16$ in steps of $1$, then quadratically increasing from $25$ to $10000$, then exponentially increasing to a maximum mean photon number of approximately $68000$). For every input state we record $570000$ samples. All data is streamed in the form of raw time tags to a computer, where it is then stored. 

Fig.~\ref{fig:setup}(b) shows experimental arrival-time histograms for different mean photon numbers (colored dots) fitted with our PNR model (black curves). The resulting fit is a sum of exponentially-modified Gaussian (EMG) distributions (Eq.~\ref{eqn:EMGsum}), weighted by the Poisson probabilities of the input states. We achieve excellent fits over several orders of magnitude of the arrival-time probability and across a wide range of $\bar{n}$ from 1 to $500$ per laser pulse. For larger mean photon numbers per pulse, the model deviates slightly from the experimental data in the exponential tail towards later arrival times. This can be seen from the brown points in Fig.~\ref{fig:setup}(b) for a mean photon number of roughly 713.

The mean value $\mu_n = \mu_\infty+\Delta\mu/\sqrt{n}$, where $\mu_\infty=144~\mathrm{ps}$ is an arbitrary offset delay in our setup (which can be seen as the time difference between the trigger signal and the steepest rising edge). The relevant fitting parameter $\Delta\mu$ is the difference of the one-photon peak $\mu_{n=1}$ and $\mu_{n=\infty}$, and will be detector-dependent.  If the parameter $\mu_\infty$ is not known, it can also be fitted in the model as $\mu_n = \mu_\infty+(\mu_1-\mu_\infty)/\sqrt{n}$, which will also result in accurate fits.

The jitter contributions to $\sigma_n$ (Eq.~\ref{eqn:sigma}) are as follows. The instrumental jitter $\sigma_\text{inst}=3$~ps (from time-tagger and trigger photodiode) and optical pulse duration $\sigma_\text{opt}=1$~ps are both independent of photon number. The noise jitter $\sigma_{\text{noise},n} = \sigma_\text{elec}/(\gamma_{\text{slew},1}\sqrt{n})=4.5/\sqrt{n}$~ps, where $\sigma_\text{elec}=4.9$~mV is the RMS electrical noise of the trace and $\gamma_{\text{slew},1}=1.08$~mVps$^{-1}$ is the independently-measured slew rate for the one-photon component.
The geometric jitter (see Sec.~\ref{sec:geom}) $\sigma_{\text{geom},n} = \sigma_{\text{geom},1} / n^{0.75}$, where $\sigma_{\text{geom},1}=9$~ps is empirically determined proportionality constant and the exponent 0.75 is independently determined numerically. The remaining intrinsic jitter contributions, $\sigma_\text{int}$ and $\tau$, are the fitting parameters.

\begin{figure*}[ht]
    \centering
    \includegraphics[width=1\linewidth]{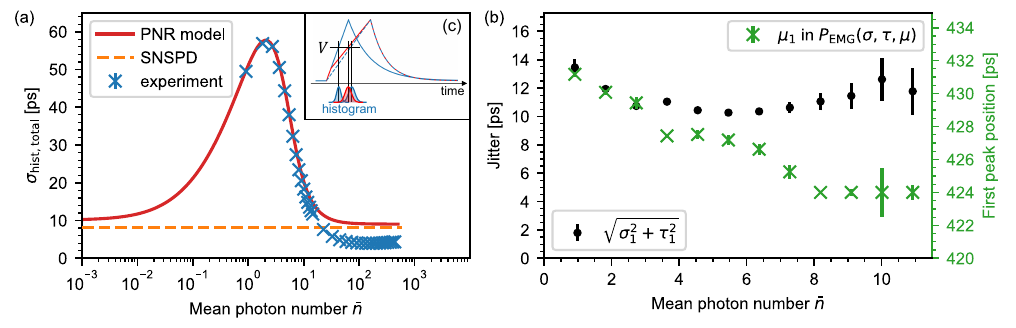}
    \caption{(a) Total width of the arrival-time histogram $\sigma_\mathrm{hist, total}$ as a function of mean photon number. Model and experiment show close agreement. The orange dashed line corresponds to the SNSPD jitter of $19~\mathrm{ps}$ (as specified by the manufacturer) when converted to units of standard deviation, which is dominant for low mean photon numbers and also fits the model. The error bars for the experimental data (not visible) are calculated using bootstrapping. (b) Best-fit jitter $\sqrt{\sigma_1^2+\tau_1^2}$ (left, black axis) and mean position $\mu_1$ of the first peak (right, green axis) from Eq.~\ref{eqn:EMGdist}. The error bars are based on the fitting uncertainty. Beyond a mean photon number of 12 the one-photon number component is non-existent (see Fig.~\ref{fig:setup}(b)). The inset (c) shows a schematic of voltage response pulses from a single domain (blue dashed line), two independent domains (blue solid line) and initially two independent domains that then grow into each other (red solid line). Overlapping domains populate earlier times in the histogram. The red histogram is not completely accurate, as the distance between domain centers can vary continuously, thus the voltage signal will follow the two-domain dynamic for different amounts of time, which will wash out the histogram.}
    \label{fig:results}
\end{figure*}

When we keep all parameters fixed for all mean photon numbers ($\Delta\mu=\mu_1-\mu_\infty=289$~ps, $\sigma_\text{int}=6$~ps, $\tau=6$~ps), Fig.~\ref{fig:results}(a) shows how the total width of the arrival-time histogram from the experimental data matches the prediction of the model. For these fixed parameters, the agreement is good until mean photon numbers up to $\sim10$, after which the effects of overlapping domains cause significant deviations. At even higher mean photon numbers, hotspot overlap effects (see Sec.~\ref{sec:sameSite}) eliminate intrinsic jitter, and the total width of the arrival time histogram is dominated by instrumentation jitter. Similar experimental data were obtained from a waveguide-integrated SNSPD~\cite{schuck2024quantum}.

\section{Discussion}\label{sec:discusssion}
While it is clear that our results agree with the model very well, we made a number of assumptions which may not be applicable in all uses. One such limitation is domain overlap, the effects of which we see at high mean photon numbers (see Fig.~\ref{fig:results}(a)). Beyond a mean photon number of approximately 20 photons per pulse, the total width of the arrival-time histogram falls significantly below the jitter of the detector as specified by the manufacturer. Equations~\ref{eqn:PNRdep} show that in this regime, the jitter due to noise and geometry can be neglected; all that remains is a residual photon-number dependent jitter in $\sigma_{\text{int},n}$ and $\tau_n$, which were assumed photon-number independent in the model. 

In scenarios of multi-photon absorption at independent sites, intrinsic jitter is influenced by the photon number due to the following: each absorbed photon initiates a delay in the emergence of a resistive domain, according to the local fluctuations, which are not correlated. The shortest delay determines the voltage pulse arrival time similar to the effect of delayed multi-photon absorption described in Sec.~\ref{sec:delayed}. As the number of photons increases, the intrinsic jitter decreases for the same reason as for multi-photon geometric jitter in Sec.~\ref{sec:geom}). Statistical analysis showed that for the studied detector this effect is small and can be neglected.
We therefore attribute the change in $\sigma_{\text{int},n}$ and $\tau_n$ to the effects of domain overlap.

\subsection{Limitations due to domain overlap}
The one-photon peak in the experimental arrival-time histograms (first peak from the right in Fig.~\ref{fig:setup}(b)), attributed to single-photon detection, is the most pronounced and provides insights into domain overlap. Fitting the one-photon peaks for a set of $\bar{n}$ from 1 to 13 with EMG distributions shows that the total jitter $\sqrt{\sigma_1^2+\tau_1^2}$ stays broadly constant in few-photon range (see Fig.~\ref{fig:results}(b), left (black) axis), but their mean positions shift to earlier times as $\bar{n}$ increases (see Fig.~\ref{fig:results}(b), right (green) axis). This shift arises because domain overlap (due to absorption of two photons in close proximity, see Sec.~\ref{sec:overlap}) introduces earlier events into the histogram. In general this leads to an overestimation of low photon-number contributions and an underestimation of $\bar{n}$.
While increasing the nanowire length can reduce the probability of domain overlap, adding a series inductor or shunt resistor would have the opposite effect, slowing current dynamics and increasing the effective domain size.

Simulations of the electrothermal model show that the voltage rise time is faster for two non-overlapping resistive domains~\cite{yang2007modeling}, producing a second peak in the histogram. However, when domains grow and overlap, the voltage response pulse initially follows the two-domain dynamics but transitions to one-domain behavior during overlap (see Fig.~\ref{fig:results}(c) for a schematic). These overlapping events populate earlier times in the first peak, shifting its mean value to earlier times as $\bar{n}$ increases, as observed. This process is of continuous nature, as the distance between two initially non-overlapping domains varies continuously. Thus the time at which the domains grow into each other and the voltage transitioning to the one-domain dynamics is also continuous.

Due to the much smaller size of hotspots compared to domain sizes, hotspot overlap becomes significant only at higher $\bar{n}$. While overlapping hotspots do not affect the rise time, they do reduce the voltage response pulse arrival time, populating earlier times in the histogram (see Fig.~\ref{fig:jitterSchematic})(a)(ii)). This shift is more pronounced for SNSPDs which operate at non-saturated detection efficiency as a function of current, which might result in an overestimation of high photon-number states.

\subsection{Implications for detector engineering} \label{sec:detEngi}
First of all it is clear that this method can only work for optical pulse durations $\sigma_\text{opt}$ below characteristic times dependent on the superconducting material and substrate properties. Below the hotspot formation time (typically below $\sim5$~ps for NbTiN~\cite{sidorova2021magnetoconductance}, $\sim15$~ps for WSi~\cite{sidorova2021magnetoconductance}, and $\sim12-20$~ps for MoSi (unpublished)) one can operate in the regime of independent photon absorption (Sec.~\ref{sec:independent}). Beyond these times, up to the resistive domain lifetime ($50$~ps~\cite{sidorova2024low} for NbTiN, $\sim800$~ps for WSi~\cite{sidorova2018nonbolometric} and $\sim70-130$~ps for MoSi (unpublished)), we operate in the regime of delayed photon absorption in independent sites (Sec.~\ref{sec:delayed}), where the dynamics is more complex but PNR signatures may still be resolvable.

Inspection of Eqs.~\ref{eqn:PNRdep} show that to optimize PNR efficiency, {i.e.,} distinguishability of different photon numbers, one should maximize $\Delta\mu$ and minimize the jitter contributions. Furthermore, one should use a large-area detector to maximize the number of independent domains, thereby eliminating the effect of domain overlap, albeit at the cost of increased geometric jitter. However, the geometric jitter drops quickly with photon number, and is significantly smaller than intrinsic jitter for micron-wide detectors, meaning this cost may be small. From Eq.~\ref{eqn:PNRdep}a, maximizing $\Delta\mu$ means increasing the rise time $t_\text{rise}$. This is achieved by increasing the nanowire's kinetic inductance $L_k$, since $t_{\mathrm{rise}} \propto L_k$. However, this also increases the ($1/\mathrm{e}$) voltage falling time $t_{\mathrm{fall}} = L_k/50\Omega$ and, therefore, the detector's reset time (dead time), which is defined as the time at which the voltage with amplitude $A$ drops below the noise level $N$, $t_{\mathrm{reset}}=t_{\mathrm{fall}}\log(N/A)$. These trade-offs must be carefully considered when optimizing the detector's design for PNR, as has been noted in Refs.~\cite{kong2024large,jaha2024kinetic}. 

Increasing the detector's rise time also increases the variance due to noise following Eq.~\ref{eqn:PNRdep}d. Reducing the electrical noise by using a low-noise cryogenic amplifier has become a well-established technique in the community, therefore enhancing PNR (and which has also been noted in Ref.~\cite{los2024high}). Moreover, increasing the rise time relaxes the bandwidth requirements since $\mathrm{BW} = 0.35/t_{\mathrm{rise}}$. Furthermore, increasing the amplitude, which means maximizing the bias current, can also help. The combination of large bias currents and large area detectors is readily achieved by widening the nanowire, {i.e.} using micron-wide wires~\cite{korneeva2018optical,charaev2020large,chiles2020superconducting}. With this approach, Kong et al.\cite{kong2024large} achieved PNR up to 10 photons, sacrificing the reset time.

It is also apparent from Eqs.~\ref{eqn:PNRdep}d and~\ref{eqn:PNRdep}e that noise and geometric jitter can be effectively removed when using large number of photons per pulse due to the inverse proportionality on $n$. This has practical implications in laser ranging: higher photon fluxes improve resolution, as multi-photon events reduce the timing jitter (due to hotspot overlap) and sharpen the histogram peak.

\vspace*{5mm}

\section{Conclusion}
In this work, we have provided a quantitative model of photon-number dependent traces of SNSPDs which includes both the arrival time and photon-number dependent jitter.
Our model accurately predicts the arrival-time probability histogram across several orders of magnitude for a broad range of mean numbers of incident photons. Implementation of the model will enable the design of detectors with maximized PNR capability. While these results are certainly promising, there are additional effects which can be incorporated into future modeling, such as double-ended or differential read-out of the SNSPDs (eliminating geometric jitter), or the use of impedance matching tapers~\cite{zhu2019superconducting}. 

\section*{Acknowledgment}
Partially funded by the European Union (ERC, QuESADILLA, 101042399). Views and opinions expressed are however those of the author(s) only and do not necessarily reflect those of the European Union or the European Research Council Executive Agency. Neither the European Union nor the granting authority can be held responsible for them. This work has received funding from the German Ministry of Education and Research within the PhoQuant project (grant number 13N16103). F.S. is part of the Max Planck School of Photonics supported by the German Federal Ministry of Education and Research (BMBF), the Max Planck Society, and the Fraunhofer Society.

\section*{Data Availability Statement}
The data that support the findings of this study are openly available in Zenodo at \url{https://doi.org/10.5281/zenodo.14888684}.

\bibliography{references}
\end{document}